\documentclass{jpsj-suppl}
\usepackage{txfonts} 
\usepackage{bm}

\title{Relationship between single-particle excitation and spin excitation at the Mott Transition}

\author{Masanori \textsc{Kohno}}

\inst{International Center for Materials Nanoarchitectonics (WPI-MANA), National Institute for Materials Science, Tsukuba 305-0044, Japan}

\email{KOHNO.Masanori@nims.go.jp}

\recdate{September 30, 2013}

\abst{An intuitive interpretation of the relationship between the dispersion relation of the single-particle excitation in a metal and 
that of the spin excitation in a Mott insulator is presented, based on the results for the one- and two-dimensional Hubbard models 
obtained by using the Bethe ansatz, dynamical density-matrix renormalization group method, and cluster perturbation theory. 
The dispersion relation of the spin excitation in the Mott insulator is naturally constructed from that of the single-particle excitation in the zero-doping limit 
in both one- and two-dimensional Hubbard models, which allows us to interpret the doping-induced states as the states that lose charge character toward the Mott transition. 
The characteristic feature of the Mott transition is contrasted with the feature of a Fermi liquid and that of the transition between a band insulator and a metal.}

\kword{Mott transition, single-particle excitation, spin excitation, Hubbard model, doping-induced states}

\begin{document}
\maketitle

\section{Introduction}
Conduction electrons in a metal usually behave as single particles carrying spin and charge. In a Mott insulator, in contrast, the low-energy charge and spin excitations are separated: 
the charge excitation has an energy gap whereas the spin excitation is usually gapless. 
A fundamental issue of the Mott transition is how electrons carrying spin and charge in a metal change into those exhibiting spin-charge separation in a Mott insulator. 
Recently, an answer to this issue has been obtained in the one-dimensional (1D) and two-dimensional (2D) Hubbard models in Refs. \citen{Kohno1DHub} and \citen{Kohno2DHub}. 
Namely, the Mott transition is characterized by freezing of the charge degrees of freedom in a single-particle excitation that leads continuously to the spin excitation of the Mott insulator. 
In this paper, to make the nature of the Mott transition more understandable, an intuitive interpretation of the relationship between the single-particle excitation in a metal and the spin excitation in a Mott insulator is presented. 
\section{Models and methods}
\label{sec:ModelsMethods} 
We consider the 1D and 2D Hubbard models defined by the following Hamiltonian. 
\vspace{-0.26mm}\hspace{-1mm}
$$
{\cal H}=-t\sum_{\langle i,j\rangle,\sigma}\left(c^\dagger_{i\sigma}c_{j\sigma}+\mbox{H.c.}\right)+U\sum_in_{i\uparrow}n_{i\downarrow}-\mu\sum_{i,\sigma}n_{i\sigma},
$$
where $c_{i\sigma}$ and $n_{i\sigma}$ denote the annihilation and number operators, respectively, of an electron at site $i$ with spin $\sigma$. 
The notation $\langle i,j\rangle$ indicates that sites $i$ and $j$ are nearest neighbors on a chain for the 1D Hubbard model and on a square lattice for the 2D Hubbard model. We consider the case for $t>0$ and $U>0$. 
The doping concentration $\delta$ is defined as $\delta=1-n$, where $n$ denotes the density of electrons. At half-filling ($\delta=0$), the system becomes a Mott insulator for $U>0$ \cite{LiebWu,HirschHub}, 
whose low-energy properties in the large-$U/t$ regime are effectively described by the Heisenberg model defined by the Hamiltonian ${\cal H}_{\rm spin}=J\sum_{\langle i,j\rangle}{\bm S}_i\cdot{\bm S}_j$, 
where ${\bm S}_i$ denotes the spin-1/2 operator at site $i$ and $J=4t^2/U$ \cite{AndersonSE}. The notation $\langle i,j\rangle$ indicates that sites $i$ and $j$ are nearest neighbors on a chain for the 1D Heisenberg model and on a square lattice for the 2D Heisenberg model. 
\par
We study the single-particle spectral function $A({\bm k},\omega)$ at zero temperature defined as follows. 
\vspace{-1mm}\hspace{-1mm}
\begin{equation}
A({\bm k},\omega)=\left\{
\begin{array}{rl}
\sum_l\left|\langle l|c^\dagger_{{\bm k}\uparrow}|\mbox{GS}\rangle\right|^2\delta\left(\omega-\varepsilon_l\right)&\quad\mbox{for}\quad\omega>0,\\
\sum_l\left|\langle l|c_{{\bm k}\downarrow}|\mbox{GS}\rangle\right|^2\delta\left(\omega+\varepsilon_l\right)&\quad\mbox{for}\quad\omega<0, 
\end{array}\right.
\label{eq:Akw}
\end{equation}
where $c_{{\bm k}\sigma}$ denotes the annihilation operator of an electron with momentum ${\bm k}$ and spin $\sigma$. 
Here, $\varepsilon_l$ denotes the excitation energy of the eigenstate $|l\rangle$ from the ground state $|\mbox{GS}\rangle$. 
The spectral function $A({\bm k},\omega)$ can also be expressed as $A({\bm k},\omega)=-\mbox{Im}G({\bm k},\omega)/\pi$, 
where $G({\bm k},\omega)$ denotes the retarded single-particle Green function \cite{ImadaRMP}. 
The spectral function $A({\bm k},\omega)$ can be probed by using angle-resolved photoemission spectroscopy \cite{ShenRMP}. 
\par
To calculate $A({\bm k},\omega)$, we employ the dynamical density-matrix renormalization group (DDMRG) method \cite{DDMRG} and cluster perturbation theory (CPT) \cite{CPTPRB,CPTPRL} for the 1D and 2D Hubbard models, respectively. 
In the DDMRG method, 120 eigenstates of the density matrix are kept for a 60-site chain \cite{Kohno1DHub}. 
In CPT, the Green function is obtained by connecting cluster Green functions calculated by exact diagonalization through the first-order single-particle hopping process without assuming long-range order. 
Since the hopping between clusters is approximated as the first-order single-particle process (which is exact at $U=0$ \cite{CPTPRB,CPTPRL}), 
CPT tends to enhance single-particle behavior like the quasiparticle in a Fermi liquid or rigid band. 
In the CPT calculation, we use cluster Green functions in ($4\times4$)-site clusters which preserve rotational and inversion symmetries of the square lattice \cite{Kohno2DHub}. 
Gaussian broadening (standard deviation $\sigma=0.1t$) is used for the spectral function \cite{Kohno1DHub,Kohno2DHub}. 
For the 1D Hubbard model, the Bethe ansatz \cite{LiebWu} is used to investigate the natures of  the characteristic modes \cite{Kohno1DHub}. 
\section{Relationship between single-particle excitation and spin excitation} 
\label{sec:RelationToSpin}
\begin{figure}
\includegraphics[width=15.5cm]{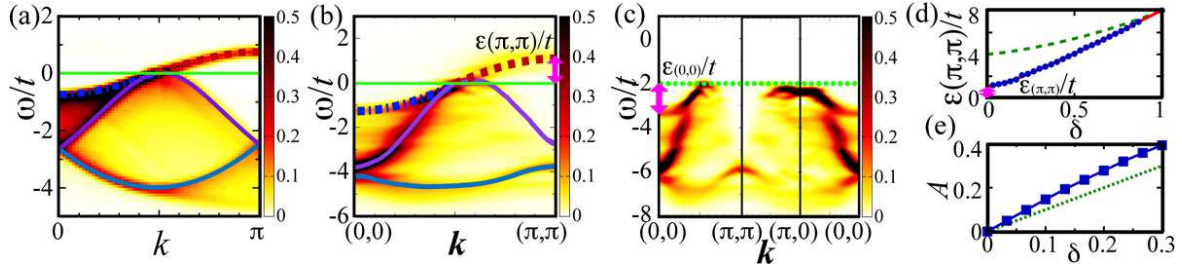}
\caption{Spectral properties of the 1D and 2D Hubbard models in the LHB for $U/t=8$. The results for 1D [(a)] are taken from Ref. \citen{Kohno1DHub} and those for 2D obtained using CPT [(b)--(e)] are from Ref. \citen{Kohno2DHub}. 
(a) $A(k,\omega)t$ of the 1D Hubbard model at $\delta\approx 0.03$ obtained using the DDMRG method \cite{Kohno1DHub}. 
The curves show the dispersion relations obtained using the Bethe ansatz \cite{Kohno1DHub}. 
The dashed red curve indicates the upper edge of the spinon-antiholon continuum. 
The dash-dotted blue curve indicates the spinon mode. The solid curves indicate the holon modes for $\omega<0$ and the antiholon mode for $\omega>0$. 
The straight solid green lines in (a) and (b) represent $\omega=0$. 
(b) $A({\bm k},\omega)t$ of the 2D Hubbard model in the $(0,0)$--$(\pi,\pi)$ direction at $\delta=0.03$ \cite{Kohno2DHub}. 
The curves indicate the modes originating from the 1D modes with the same line types as in (a). 
The pink arrow indicates the energy of the mode for $\omega>0$ at $(\pi,\pi)$ [$\varepsilon(\pi,\pi)$]. 
(c) $A({\bm k},\omega)t$ of the 2D Hubbard model at $\delta=0$ \cite{Kohno2DHub}. 
The dotted green line indicates the $\omega$ value at the top of the LHB. 
The pink arrow indicates the bandwidth of the mode originating from the 1D spinon mode [$\varepsilon_{(0,0)}$]. 
(d) Doping dependence of $\varepsilon(\pi,\pi)/t$ in the 2D Hubbard model for $U/t=8$ (blue circles with solid red line) \cite{Kohno2DHub}. 
The pink arrow indicates the extrapolated value of $\varepsilon(\pi,\pi)$ to the $\delta\rightarrow 0$ limit [$\varepsilon_{(\pi,\pi)}$]. 
The dashed green curve represents $\varepsilon(\pi,\pi)/t$ for $U=0$. 
(e) Spectral weight $A$ for $\omega>0$ in the LHB of the 2D Hubbard model (blue squares with solid line) \cite{Kohno2DHub}. 
The dotted green line indicates that for $t=0$ \cite{Eskes}.}
\label{fig:Identification}
\end{figure}

The origins of the characteristic modes in the $(0,0)$--$(\pi,\pi)$ direction in the 2D Hubbard model [Fig. \ref{fig:Identification}(b)] can be traced back to those of the 1D Hubbard model [Fig. \ref{fig:Identification}(a)], 
by considering the spectral-weight shift caused by interchain hopping \cite{Kohno2DHub}. 
The most characteristic feature of the Mott transition is the behavior of the dispersing mode for $\omega>0$ in the lower Hubbard band  (LHB) [Figs. \ref{fig:Identification}(a) and \ref{fig:Identification}(b), dashed red curves]. 
This mode corresponds to the states referred to as doping-induced states or in-gap states \cite{Eskes,DagottoDOS,DagottoRMP,PreussQP,PreussPG,PhillipsRMP,SakaiImada,ImadaCofermion}. 
In both 1D and 2D Hubbard models, its spectral weight gradually disappears toward Mott transition [Fig. \ref{fig:Identification}(e)], 
while the dispersion relation remains dispersing even in the $\delta\rightarrow 0$ limit [Fig. \ref{fig:Identification}(d)] \cite{Kohno1DHub,Kohno2DHub}. 
In addition, its dispersion relation, as well as that of the mode for $\omega<0$ in the low-$|\omega|$ regime [Figs. \ref{fig:Identification}(a) and \ref{fig:Identification}(b), dash-dotted blue curves], 
is directly related to the dispersion relation of the spin excitation of the Mott insulator as shown below \cite{Kohno1DHub,Kohno2DHub}. 
\par
\begin{figure}
\includegraphics[width=15.5cm]{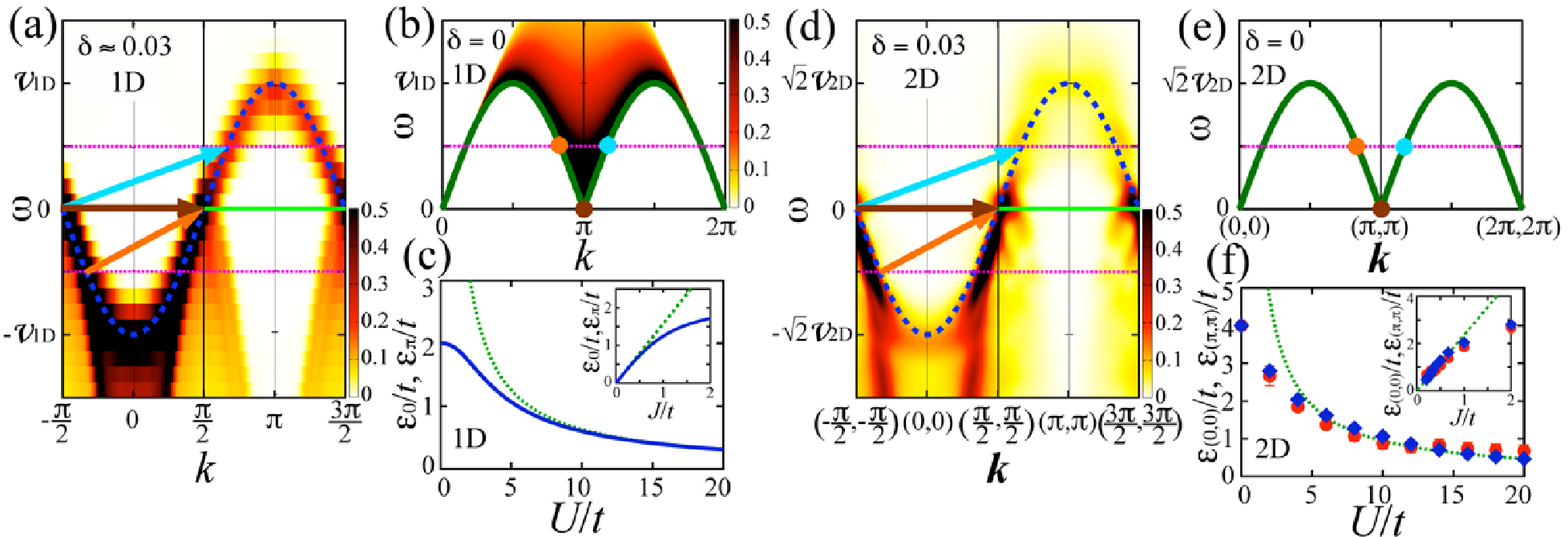}
\caption{Relationship between the dispersion relation of single-particle excitation near the Mott transition and that of spin excitation in the Mott insulator. 
(a) $A(k,\omega)t$ of the 1D Hubbard model for $U/t=8$ at $\delta\approx0.03$ obtained using the DDMRG method, taken from Ref. \citen{Kohno1DHub}. 
The dashed blue curve indicates $\varepsilon_{\rm 1D}(k)$ [Eq.~(\ref{eq:1Dspinon})]~\cite{Kohno1DHub}~where the spin-wave velocity of the 1D Heisenberg model $v_{\rm 1D}$($=\pi J/2$ \cite{desCloizeaux}) for $U/t=8$ is used ($J=4t^2/U$). 
The solid green lines in (a) and (d) represent $\omega=0$. 
(b) Spin-wave dispersion relation of the 1D Heisenberg model (solid green curve) \cite{desCloizeaux}. 
The circles correspond to excitations shown by the arrows with the same colors as~in~(a). 
The intensity plot shows the dynamical structure factor $S(k,\omega)J$ of two-spinon excitations obtained using the Bethe ansatz \cite{Biegel}. 
(c) Bandwidth of the spinon mode at $\delta=0$ [$\varepsilon_0$] (solid blue curve) and 
energy of the mode for $\omega>0$ in the LHB at $k=\pi$ in the $\delta\rightarrow 0$ limit [$\varepsilon_{\pi}$] (solid blue curve) 
in single-particle excitation of the 1D Hubbard model obtained using the Bethe ansatz \cite{Kohno1DHub}, taken from Ref. \citen{Kohno2DHub}. 
The dotted green line indicates $v_{\rm 1D}/t[=\pi J/(2t)]$. The inset shows the $J/t$-dependence. 
(d) $A({\bm k},\omega)t$ of the 2D Hubbard model in the $(0,0)$--$(\pi,\pi)$ direction for $U/t=8$ at $\delta= 0.03$ obtained using CPT, taken from Ref. \citen{Kohno2DHub}. 
The dashed blue curve indicates $\varepsilon_{\rm 2D}(k,k)$ [Eq. (\ref{eq:2Dspinon})] \cite{Kohno2DHub} 
where the spin-wave velocity of the 2D Heisenberg model $v_{\rm 2D}$($=1.18\sqrt 2J$ \cite{Singh}) for $U/t=8$ is used. 
(e) The same as (b) but for the 2D Heisenberg model in the $(0,0)$--$(\pi,\pi)$ direction \cite{AndersonSW}. 
The circles correspond to the arrows with the same colors as in (d). 
(f) Bandwidth of the mode originating from the 1D spinon mode at $\delta=0$ [$\varepsilon_{(0,0)}$] [Fig. \ref{fig:Identification}(c)] (blue diamonds) and 
energy of the mode for $\omega>0$ in the LHB at $(\pi,\pi)$ in the $\delta\rightarrow 0$ limit [$\varepsilon_{(\pi,\pi)}$] [Fig. \ref{fig:Identification}(d)] (red circles) 
in single-particle excitation of the 2D Hubbard model obtained using CPT, taken from Ref. \citen{Kohno2DHub}. 
The dotted green line indicates $\sqrt{2}v_{\rm 2D}/t(=2.36J/t)$. The inset shows the $J/t$-dependence.}
\label{fig:spin}
\end{figure}

The results obtained by using CPT for the 2D Hubbard model \cite{Kohno2DHub} have indicated that the energy of the mode for $\omega>0$ in the LHB 
[Fig. \ref{fig:Identification}(b), dashed red curve] at $(\pi,\pi)$ in the $\delta\rightarrow0$ limit, 
denoted as $\varepsilon_{(\pi,\pi)}$ [Fig. \ref{fig:Identification} (d)], and the bandwidth of the mode primarily originating from the 1D spinon mode for $\omega<0$ at $\delta=0$, 
denoted as $\varepsilon_{(0,0)}$ [Fig. \ref{fig:Identification} (c)], behave as $\sqrt 2 v_{\rm 2D}$ in the large-$U/t$ regime, 
where $v_{\rm 2D}$ denotes the spin-wave velocity of the 2D Heisenberg model ($v_{\rm 2D}\approx1.18\sqrt 2 J$ \cite{Singh}) with $J=4t^2/U$ [Fig. \ref{fig:spin}(f)]. 
In the 1D Hubbard model, $\varepsilon_{0}$ and $\varepsilon_{\pi}$, defined in a similar fashion to $\varepsilon_{(0,0)}$ and $\varepsilon_{(\pi,\pi)}$, 
reduce to $v_{\rm 1D}$ in the large-$U/t$ limit \cite{Kohno1DHub,EsslerBook,TakahashiBook}, 
where $v_{\rm 1D}$ denotes the spin-wave velocity of the 1D Heisenberg model ($v_{\rm 1D}=\pi J/2$ \cite{desCloizeaux}) with $J=4t^2/U$ [Fig. \ref{fig:spin}(c)]. 
This implies that the mode of the single-particle excitation in the low-$|\omega|$ regime for $\omega>0$, as well as that for $\omega<0$, 
leads continuously to the spin excitation of the Mott insulator \cite{Kohno1DHub,Kohno2DHub}. 
Below, we interpret this feature more intuitively. 
\par
In 1D, the dispersion relation of the spin excitation (two-spinon excitation) of the Heisenberg model has been obtained by using exact solutions \cite{Yamada,Muller,TakahashiBook}, 
which can be expressed as 
\vspace{-1mm}\hspace{-1mm}
\begin{eqnarray}
\label{eq:1Dspin}
E_{\rm 2spinon}(k)&=&\varepsilon_{\rm 1D}(k_2)-\varepsilon_{\rm 1D}(k_1), \quad k=k_2-k_1,\\
\varepsilon_{\rm 1D}(k)&=&-v_{\rm 1D}\cos k, 
\label{eq:1Dspinon}
\end{eqnarray}
where $|k_1|\le\pi/2$ and $|k_2-\pi|\le\pi/2$ for $0\le k\le2\pi$ \cite{Yamada}. Here, the spin-wave velocity has been obtained to be $v_{\rm 1D}=\pi J/2$ \cite{desCloizeaux}. 
The dominant part is the lower edge of the two-spinon continuum [Fig. \ref{fig:spin}(b)] \cite{Muller,Biegel}, 
whose dispersion relation is obtained by setting $k_1=-\pi/2$ or $k_2=\pi/2$ as \cite{desCloizeaux} 
\vspace{-1mm}\hspace{-1mm}
\begin{equation}
E(k)=|\varepsilon_{\rm 1D}(k-\pi/2)|=v_{\rm 1D}|\sin k|.
\label{eq:1DspinDP}
\end{equation}
By noting that the single-particle dispersion relation of the 1D Hubbard model reduces to $\varepsilon_{\rm 1D}(k)$ 
as $\delta\rightarrow0$ and $U/t\rightarrow\infty$ [Fig. \ref{fig:spin}(a)] \cite{Kohno1DHub,TakahashiBook,EsslerBook}, 
the spin-wave dispersion relation [Eq. (\ref{eq:1DspinDP})] can be interpreted as follows. 
The gapless point of the spin-flip excitation at $k$=$\pi$ [Fig. \ref{fig:spin}(b), brown circle] is due~to the zero-energy excitation 
obtained by removing a spin-$\downarrow$ quasiparticle at $k=-\pi/2$ and adding~a spin-$\uparrow$ quasiparticle at $k=\pi/2$, as indicated by the brown arrow in Fig.~\ref{fig:spin}(a). 
The spin-wave dispersion relation [Eq. (\ref{eq:1DspinDP}); Fig. \ref{fig:spin}(b), solid green curve] is obtained 
by shifting either the starting or end point of the arrow along the single-particle dispersion relation [Eq. (\ref{eq:1Dspinon}); Fig. \ref{fig:spin}(a),~dashed~blue~curve], 
as~shown by the examples of the orange and light blue arrows in Fig. \ref{fig:spin}(a) \cite{Yamada}. 
Reflecting the symmetry of spin excitation with respect to $k=\pi$ [Eq. (\ref{eq:1Dspin}); Fig. \ref{fig:spin}(b)], 
the single-particle dispersion relation in the $\delta\rightarrow0$ limit is symmetric 
with respect to the gapless point [Eq. (\ref{eq:1Dspinon}); Fig. \ref{fig:spin}(a), dashed blue curve] \cite{Kohno1DHub}. 
\par
This feature also appears in the $(0,0)$--$(\pi,\pi)$ direction in the 2D Hubbard model \cite{Kohno2DHub}. 
In spin-wave theory \cite{AndersonSW}, the dispersion relation of the spin-wave mode in the 2D Heisenberg model is obtained as 
\vspace{-1mm}\hspace{-1mm}
\begin{equation}
E({\bm k})=\sqrt{2}v_{\rm 2D}\sqrt{1-\left(\frac{\cos k_x+\cos k_y}{2}\right)^2}, 
\label{eq:spinwave}
\end{equation}
\vspace{-0.2mm}\hspace{-1mm}
where the spin-wave velocity has been estimated as $v_{\rm 2D}\approx1.18\sqrt2 J$ \cite{Singh}. 
For $k_x=k_y(=k)$, 
\vspace{-1mm}\hspace{-1mm}
\begin{equation}
E(k,k)=\sqrt{2}v_{\rm 2D}|\sin k|, 
\label{eq:spinwavediag}
\end{equation}
which can also be expressed as 
\vspace{-1mm}\hspace{-1mm}
\begin{eqnarray}
\label{eq:2Dspin}
E(k,k)&=&|\varepsilon_{\rm 2D}(k-\pi/2,k-\pi/2)|,\\
\varepsilon_{\rm 2D}(k,k)&=&-\sqrt2 v_{\rm 2D}\cos k.
\label{eq:2Dspinon}
\end{eqnarray}
The single-particle dispersion relation of the mode originating from the 1D spinon mode and that of the mode for $\omega>0$ in the LHB in the 2D Hubbard model in the large-$U/t$ regime 
have been shown to behave as $\varepsilon_{\rm 2D}(k,k)$ in the $\delta\rightarrow0$ limit [Fig. \ref{fig:spin}(d)] \cite{Kohno2DHub}. 
By following the argument in 1D, the gapless point of the spin-flip excitation at $(\pi,\pi)$ [Fig. \ref{fig:spin}(e), brown circle] can be interpreted as 
the zero-energy excitation 
obtained by removing a spin-$\downarrow$ quasiparticle at $(-\pi/2,-\pi/2)$ and adding~a spin-$\uparrow$ quasiparticle at $(\pi/2,\pi/2)$, 
as indicated by the brown arrow in Fig. \ref{fig:spin}(d). 
The spin-wave dispersion relation [Eq. (\ref{eq:spinwavediag}); Fig. \ref{fig:spin}(e), solid green curve] is obtained by shifting either the starting or end point of the arrow 
along the single-particle dispersion relation [Eq. (\ref{eq:2Dspinon}); Fig. \ref{fig:spin}(d), dashed blue curve], as shown by the examples of the orange and light blue arrows in Fig. \ref{fig:spin}(d). 
Reflecting the symmetry of spin excitation with respect to $(\pi,\pi)$ at $\delta=0$ [Eq. (\ref{eq:spinwavediag}); Fig. \ref{fig:spin}(e)], 
the single-particle dispersion relation near $\delta=0$ is almost symmetric 
with respect to the gapless point in the $(0,0)$--$(\pi,\pi)$ direction~[Fig.~\ref{fig:spin}(d)] \cite{Kohno2DHub}. 
\par
Thus, in both 1D and 2D cases, the dispersion relation of the dominant part of the spin excitation is obtained 
by shifting either the starting or end point of the arrow along the single-particle dispersion relation in Figs. \ref{fig:spin}(a) and \ref{fig:spin}(d) 
with the other anchored at the gapless point. 
By extending the argument, we can interpret the behavior of the spin excitation in terms of the single-particle excitation. 
The behavior of the single spin-wave mode carrying most of the spectral weights in 2D can be interpreted by considering that the spin-wave mode 
constructed as described above [Eq. (\ref{eq:2Dspin}); Figs. \ref{fig:spin}(d) and \ref{fig:spin}(e)] carries most of the spectral weights. 
In 1D, considerable spectral weights of the spin excitation spread over the two-spinon continuum [Fig. \ref{fig:spin}(b)] \cite{Muller,Biegel}. 
This behavior can be interpreted by considering that the spin excitation obtained 
by shifting both starting and end points of the arrow along the single-particle dispersion relation in Fig. \ref{fig:spin}(a) [Eq. (\ref{eq:1Dspin})] \cite{Yamada} carries considerable spectral weights. 
\par
The continuous evolution to the spin excitation of the Mott insulator is consistent with the scaling behavior of spin correlations toward the Mott transition \cite{ImadaRMP,Furukawa,KohnotJ}. 
The gapless nature of the single-particle excitation in the $(0,0)$--$(\pi,\pi)$ direction [Eq. (\ref{eq:2Dspinon}); Fig. \ref{fig:spin}(d)] in the 2D Hubbard model \cite{Kohno2DHub} 
is compatible with that of the cuprate high-temperature superconductors having the order parameter with $d_{x^2-y^2}$-wave symmetry \cite{ShenRMP}, 
as well as the gapless spin-wave mode created from the antiferromagnetic long-range order in the Mott insulator \cite{AndersonSW,LCO_SW}. 
Although some recent studies for the 2D Hubbard model suggest that there is an energy gap between the doping-induced states and the states below them \cite{PhillipsRMP,SakaiImada,ImadaCofermion}, 
the present results relating the doping-induced states to the spin-wave mode indicate that such an energy gap does not exist in the 2D Hubbard model \cite{Kohno2DHub} as in the 1D case \cite{Kohno1DHub}. 
\par
The values of $\varepsilon_{(0,0)}$ in the large-$U/t$ regime in the 2D Hubbard model \cite{Kohno2DHub} are consistent 
with those estimated in the single-hole excitation in the 2D $t$-$J$ model at $\delta=0$ \cite{DagottoRMP,Poilblanc1h,Dagotto1h}. 
Although $\varepsilon_{(0,0)}$ has been considered in the literature to behave according to a power law 
as a function of $J$ with an exponent of 0.7 $\lesssim\alpha\lesssim 1$ \cite{DagottoRMP,Poilblanc1h,Dagotto1h}, 
the present analyses based on the features derived from the exact 1D results \cite{Kohno1DHub} indicate that the value of $\varepsilon_{(0,0)}$ can be interpreted as $\sqrt 2 v_{\rm 2D}$ \cite{Kohno2DHub}. 
\par
In the extremely large-$U/t$ regime, ferromagnetic fluctuations arising from Nagaoka ferromagnetism \cite{Nagaoka,PutikkaFM,ZhangFM,KohnoUinfLadder} 
might affect the single-particle excitation as well as the spin excitation near the Mott transition, an outcome that is beyond the scope of the present study. 
\section{Summary and discussion}
In this paper, the relationship between the single-particle excitation in the $\delta\rightarrow 0$ limit and the spin excitation at $\delta=0$ has been clarified. 
Near the Mott transition of the 1D and 2D Hubbard models, the mode of the low-$|\omega|$ single-particle excitation remains dispersing even in the $\delta\rightarrow 0$ limit, 
and its dispersion relation is directly related to that of the spin excitation of the Mott insulator \cite{Kohno1DHub,Kohno2DHub}. 
Due to this feature, the reduction in spectral weight from the dispersing mode for $\omega>0$ in the LHB can be interpreted as a loss of charge character toward the Mott transition \cite{Kohno1DHub,Kohno2DHub}. 
This is the nature of the Mott transition, which reflects the spin-charge separation (non-zero charge gap and gapless spin excitation) in the Mott insulating phase \cite{Kohno1DHub,Kohno2DHub}. 
This characteristic feature of the Mott transition is contrasted with the feature of the transition between a band insulator and a metal: 
doping of a band insulator neither changes the band structure nor induces an additional dispersing mode, 
because electrons continue to behave as single particles before and after the transition without exhibiting the spin-charge separation. 
\par
In a Fermi liquid as the effective mass $m^*\rightarrow\infty$, the effective bandwidth of the coherent part $\varepsilon^*$ shrinks as $\varepsilon^*\propto1/m^*\rightarrow 0$ \cite{ImadaRMP}. 
Thus, the characteristic feature of the Mott transition (dispersing mode losing the spectral weight) cannot be interpreted as a property of the coherent part. 
If one dares to interpret this feature in the Fermi liquid theory, it should be regarded as a contribution from the incoherent part. 
Note, however, that the mode for $\omega>0$ in the LHB seems to be continuously deformed from the effectively non-interacting mode (coherent part) in the low-electron-density limit~\cite{Kohno2DHub}. 
\par
We can confirm that the present results for the 2D Hubbard model \cite{Kohno2DHub} are not artifacts of the CPT approximation, by considering the nature of the approximation. 
CPT does not assume antiferromagnetic long-range order but rather tends to enhance paramagnetic behavior. 
Nevertheless, the CPT results show that the dispersion relation of the single-particle excitation in the metallic phase leads continuously 
to that of the spin-wave mode created from the antiferromagnetic long-range order in the Mott insulating phase \cite{Kohno2DHub}. Thus, this feature is not an artifact of the CPT approximation. 
In addition, ($4\times4$)-site clusters that preserve rotational and inversion symmetries of the square lattice are used in the present study, and CPT rather tends to enhance single-particle behavior. 
Nevertheless, the CPT results show that multiple modes similar to those of the 1D model appear \cite{Kohno2DHub}. Thus, this feature is not an artifact of the CPT approximation, either. 
These features should be intrinsic to the model.
\par
Although the Mott transition can be a first-order transition in real materials due to various factors that are not taken into account in the present study, 
the signatures for the characteristic features of the Mott transition obtained in the 1D and 2D Hubbard models \cite{Kohno1DHub,Kohno2DHub}, 
which are models that have been simplified and idealized to extract the essence of the Mott transition, are expected to generally persist near the Mott transition. 
\section*{Acknowledgements}
This work was supported by KAKENHI (No. 22014015 and 23540428) and the World Premier International Research Center Initiative (WPI), MEXT, Japan. 
The numerical calculations were partly performed on the supercomputer at the National Institute for Materials Science.

\end{document}